\documentclass[floatfix,showpacs,amsmath,amssymb,letterpaper,groupaddresses,superscriptaddress]{article}
\usepackage[export]{adjustbox}
\usepackage{latexsym}
\usepackage{epsfig}
\usepackage{amsthm}
\usepackage{amsmath}
\usepackage{amssymb}
\usepackage{graphicx}
\usepackage{times}
\usepackage[section]{placeins}
\usepackage{caption}
\usepackage{subcaption}
\usepackage{physics}
\usepackage{forest}
\usepackage{tikz}
\usepackage{enumitem}
\usepackage[font=footnotesize,labelfont=bf]{caption}
\usetikzlibrary{calc,backgrounds,fit,decorations.pathreplacing}  
\usetikzlibrary{arrows.meta, decorations.pathreplacing, shadows}
\usepackage[bookmarks = true, pdfpagemode = None, pdfstartview = FitH, colorlinks = true, urlcolor = blue]{hyperref}
\usepackage{qcircuit}

\usepackage{color}

\DeclareMathOperator{\cnot}{\text{CNOT}}

\newcommand{\qwmulti}[1][-1]{\ar @3{-} [0,#1]}

\def\a{\alpha}
\def\b{\beta}

\def\be{\begin{equation}}
\def\ee{\end{equation}}
\def\ba{\begin{eqnarray}}
\def\ea{\end{eqnarray}}
\def\la{\langle}
\def\ra{\rangle}
\def\a{\alpha}

\def\b{\beta}

\def\h{\hskip 1cm}

\def\lo{\longrightarrow}

\theoremstyle{definition}
\newtheorem{definition}{Definition}

\newtheorem{example}{Example}
\newtheorem*{remark}{Remark}

\theoremstyle{plain}

\begin{document}

\vspace{4cm}
\begin{center}{\Large \bf Computing on Quantum Shared Secrets for General Quantum Access Structures}\\
\vspace{2cm}

Roozbeh Bassirian$^1$, Sadra Boreiri$^{1,2}$, and Vahid Karimipour$^2$\\
\vspace{1cm} $^1$ Department of Computer Engineering, Sharif University of Technology, P.O. Box 11155-9161, Tehran, Iran.\\
$^2$ Department of Physics, Sharif University of Technology, P.O. Box 11155-9161, Tehran, Iran.\\

\end{center}
\vskip 1cm



\begin{abstract}
 Quantum secret sharing is a method for sharing a secret quantum state among a number of individuals such that certain authorized subsets of participants can recover the secret shared state by collaboration and other subsets cannot. In this paper,  we first propose a method for sharing a quantum secret in a basic $(2,3)$ threshold scheme, only by using qubits and the 7-qubit CSS code. Based on this $(2,3)$ scheme,  we propose a new $(n, n)$ scheme and we also construct a quantum secret sharing scheme for any quantum access structure by induction. Secondly, based on the techniques of performing quantum computation on 7-qubit CSS codes, we introduce a method that authorized subsets can perform universal quantum computation on this shared state, without the need for recovering it. This generalizes recent attempts for doing quantum computation on $(n, n)$ threshold schemes.
 
\end{abstract}

\vskip 0.5cm

\section{Introduction}
Secret sharing refers to procedures for distributing a secret among a group of participants, each of whom is allocated a share of the secret such that only certain qualified subsets of participants, known as authorized parties, can collaboratively reconstruct the secret. Secret sharing was first introduced independently by Adi Shamir \cite{shamir1979share} and George Blakley\cite{blakley1979safeguarding} in 1979, where both secret and shares were classical information. Quantum Secret Sharing (QSS) is a natural extension of the classical protocol to the quantum realm \cite{hillery1999quantum,karlsson1999quantum,10} where quantum mechanics provides a way for security of sharing the classical secret. This was even further extended to the case of Quantum State Sharing, in which a quantum state $|\psi\ra$ is distributed among a group of parties in such a way that it can be retrieved only by their collaboration \cite{QSS_KN,QSS_GEN}. In addition, some scholars have worked on information theoretical models for the quantum secret sharing \cite{imai2003quantum,bai2016generalized}.\\

A most interesting line of development in this subject concerns itself not only with methods of sharing and retrieving a state, but also with ways of performing  universal quantum computation on such a state by the authorized parties, who may not even need to know what the state is. This is a subject in the field of distributed computation and quantum cryptography, the latter being one of the most promising fields in quantum computation. While some of the well-established protocols have been experimentally performed \cite{1, 2, 3}, and even commercialized, new ones with different domains of applications are being proposed. Blind quantum computation \cite{4, 5}, quantum homomorphic encryption \cite{6}, sharing of classical data \cite{7,8,9} and quantum states \cite{10,QSS_KN,QSS_GEN,gordon2006generalized}, proactive quantum secret sharing \cite{qin2015proactive}, and even performing quantum computation on shared secrets \cite{HQSS_NN} are good examples of these protocols.\\

To be more specific, suppose Alice wants to encode a quantum state $|\psi\ra=\a|0\ra+\b|1\ra$ to $|\overline{\psi}\ra=\a|\overline{0}\ra+\b|\overline{1}\ra$ and shares it among $n$ participants such that only certain subsets can retrieve the state.\\

The most common and the simplest access structure, is the one denoted by $(n,n)$ where the only authorized subset is the whole set. It generalizes to the $(k,n)$ access structure,  where any subset of size $k$ can recover the data by collaboration \cite{QSS_KN}. In this so called threshold schemes, the simplest one is the $(2,3)$ scheme proposed in \cite{QSS_KN}, which uses qutrits (3-level states) and the following encoding:
	\ba\label{got}
	|\overline{0}\ra&=\frac{1}{\sqrt{3}} (|000\rangle+|111\rangle+|222\rangle) \\ 
	|\overline{1}\ra&=\frac{1}{\sqrt{3}}(|012\rangle+|120\rangle+|201\rangle) \\
	|\overline{2}\ra&=\frac{1}{\sqrt{3}}(|021\rangle+|102\rangle+|210\rangle).
	\ea
	
	It is readily seen that any two members (say 1 and 2) can retrieve the state by simply performing the operator $C_{21}C_{12}$ on their qutrits, where $C_{ij}$ is the CNOT operator which is controlled by $i$ and acts on $j$ in the form $C_{ij}|\a,\b\ra_{i,j}=|\a,\a+\b\ra_{ij}.$\\

There are many applications where the members do not have an equal level of authorization. The most general access structure is where only certain subsets of the set of receivers can retrieve the classical or the quantum data. An example of an access structure is when the whole set is $X=\{A,B,C,D\}$ and the authorized subsets are 
${\cal A}(X)=\{\{A,B,C\},\{B,C,D\}\}$. Obviously this is not a threshold scheme, since the subset $\{A,B,D\}$, although having $3$ members is not an authorized subset. \\

While there has been a lot of progress in QSS schemes for $(n,n)$ and $(k,n)$ threshold structures,  much less has been reported on these schemes for general access structures. In this respect we can mention \cite{QSS_KN, QSS_GEN}, where arbitrary quantum access structures are constructed and a relation with quantum error correcting codes has been established. We should emphasize however that these schemes use $d-$ level states for quantum state sharing, as specified in the example of equation (\ref{got}) and $d$ increases with complexity of access structure. In addition, some recent attempts to reduce the number of quantum shares to make more efficient schemes have been done \cite{fortescue2012reducing, bai2017quantum}.\\

On the other hand, quite recently it has been shown \cite{HQSS_NN} that shared and secret quantum computation is possible on an  $(n,n)$ scheme, using only qubits.  
For performing a specific computation, each party member applies a relevant operation on his or her corresponding share. They are also allowed to use ancilla qubits and public announcement of their measurement results. 
 While the desired quantum circuit is known to every party member, during the computation no  information  leaks to the un-authorized parties, neither  about  the input nor the output secret state \cite{HQSS_NN}.  \\ 

It might seem natural that performing a distributed secret computation on even a threshold access structure, let alone the general access structure, should utilize high $d-$ level states for which theoretical and experimental tools are not as developed as for qubits. The aim of our work is to fill this gap and to construct QSS schemes for general access structures using qubits, and to perform distributed quantum computation on them. \\
  
  Our method is based on general ideas of \cite{QSS_GEN} and \cite{HQSS_NN}. We use the same induction steps proposed in \cite{QSS_GEN}. However, in all the steps we use new QSS schemes based on the 7-qubit code as building blocks, and we use our purification method which makes distributed quantum computation possible for these schemes. While in \cite{QSS_GEN} it is proved that any purification method works, the steps of purification are not explicitly specified. This explicit purification is necessary if we want to do quantum computation on these shared quantum states. It is worth mentioning that using our purification method, it is also possible to do universal quantum computation on the QSS schemes proposed in \cite{QSS_GEN}.  However, it would require ancillary states for every gate. The advantage of using 7-qubit code is that it limits the usage of ancillary states to only the $\frac{\pi}{8}$-gate. \\
  
  The structure of this paper is as follows:
In Section \ref{section2}, we explain our notations and conventions.  In Section \ref{section3}, two  QSS schemes are proposed for two basic access structures which will act as building blocks of arbitrary access structures.  Section \ref{section4} shows how universal computation is possible on these basic access structures and section \ref{section5} is devoted to computation on general access structures. The paper ends with  a conclusion and outlook.

\section{Preliminaries} \label{section2}
We assume that the reader is familiar with basic concepts of error correcting codes\cite{QEC_GEN} and stabilizer formalism\cite{QEC_STAB}. In this section, we review some definitions and notations for future use.\\
In the context of secret sharing, an access structure identifies whether a group of parties should have access to a particular data. In set-theoretic concepts, an access structure marks every subset of a group as authorized or unauthorized. Thus, authorized subsets of an access structure are those subsets that are qualified to access the desired data. More formally, we have:\\

\begin{definition}
  For a given set $X$ of players, an \textbf{access structure} ${\cal A}(X)$ is a collection of authorized subsets of $X$, ${\cal A}(X) \subseteq 2^X$, where $2^X$ is the power set of $X$,
  with the monotone increasing property which is a natural property for authorized subsets. More precisely, if $S$ belongs to ${\cal A}(X)$, every superset $T$ of $S$ (i.e. any set $T$ where $S \subseteq T$) should also belong to ${\cal A}(X)$
\end{definition}

\begin{example}
We use the notation of $(k, n)$ for threshold schemes, which basically refers to an access structure in which there exist $n$ players and every subset of at least $k$ parties is authorized. For example for the $(2,3)$ threshold scheme when the whole set is $X=\{A,B,C\}$, the authorized subsets are 
${\cal A}(X)=\{\{A,B\},\{A,C\},\{B,C\},\{A,B,C\}\}$. 
\end{example}
 
For the sake of brevity  a subset like $\{A,B\}$ is denoted simply by $AB$. Thus the previous access structure is simply denoted by ${\cal A}(X)=\{AB, AC, BC, ABC\}$. To prevent cluttering of notation, in all the discussions and figures which follow, we use the same letter $A$ both for the player and for the (classical or quantum) share he or she receives in the scheme. In cases like above that a player $A$ receives multiple shares, they are denoted by subscripts i.e. $A_1$, $A_2$, etc.

\begin{definition}
	
  For a given set $X$ of players, a \textbf{quantum access structure} ${\cal A}(X)$ is an access structure on $X$ which also satisfies an extra condition imposed by  the no-cloning theorem: For every $S,T \in {\cal A}(X)$, $S \cap T \neq\varnothing$.
\end{definition}

\begin{remark}
Not all the access structures are admissible in the quantum world and there cannot exist disjoint authorized subsets in a quantum access structure. If two disjoint authorized subsets exist in ${\cal A}(X)$ then the following procedure can be used to make two disjoint copies of an arbitrary quantum state. First, apply ${\cal A}(X)$ scheme to the state, then take two disjoint authorized subsets and reconstruct two copies of the state. This contradicts the no-cloning theorem, which asserts that no operation can generate multiple copies of an unknown arbitrary quantum state \cite{NOCLON}.
\end{remark}

Hereafter, whenever we mention access structure, we mean quantum access structure.

\begin{definition}

		In every access structure, there are a number of subsets which we call the {\bf minimal authorized subsets}. Due to the monotone property of the access structure, every larger subset which contains one of these subsets or the union of them is automatically authorized and need not be written explicitly in the structure.  The notation simplifies a lot if we denote any access structure only by its minimal authorized subsets inside a bracket. An access structure ${\cal A}$ is usually indicated by ${\cal A}=\la T_1, T_2, \cdots, T_r\ra$, where $T_i$s are its minimal authorized subsets. Thus in example 1, we write  ${\cal A}(X)=\{AB, AC, BC, ABC\}=\la AB, AC, BC\ra$.
\end{definition} 

	Of special importance are the class of maximal access structures. Consider the set $X=\{A, B, C\}$, the  $(2,3)$ threshold scheme access structure ${\cal A}(X)=\{AB,AC,BC,ABC\}$, has the property that for any member of $2^X$, either itself or its complement  belong to ${\cal A}(X)$. More formally we have:

 \begin{definition} \cite{QSS_GEN} 
For a given set $X$ of players, an access structure is \textbf{maximal} and denoted by $\overline{{\cal A}}(X)$ if for every $S \subset X$, it satisfies:\\
\begin{enumerate}
\item[($i$)] If $S \in \overline{{\cal A}}(X)$, then $S^c \not \in \overline{{\cal A}}(X)$
\item[($ii$)] If $S \not \in \overline{{\cal A}}(X)$, then $S^c  \in \overline{{\cal A}}(X)$
\end{enumerate}
where $S^c = X \setminus S$. 
\end{definition}

Obviously  the threshold schemes $(n,n)$ are  not maximal.   As another example, for the set $X=\{A, B, C, E\}$, the structure $\overline{{\cal A}}=\la AE, BE, CE, ABC \ra$ is maximal while ${\cal A}=\la AE, BE, ABC \ra$ is not.  Moreover if from a maximal access structure a member is removed, the resulting structure will no longer be maximal. \footnote{More precisely, if the discarded share contains no important data, which means it is not included in any minimal authorized set, the resulting access structure is still maximal. This  means that the purification produces a redundant share, that is what we desire.}\\

  Let $X=\{A_1, A_2, A_3, \cdots\}$ be a set. The aim of quantum state sharing is to encode a qubit state $|\psi\ra=\a|0\ra+\b|1\ra$ to a multi-qubit pure state $\ket{\overline{\psi}}=\a|\overline{0}\ra_{\overline{X}}+\b|\overline{1}\ra_{\overline{X}}$ and share it to the members of $X$ such that every authorized subset in the access structure ${\cal A}(X)$ can recover the original state and no unauthorized subset can retrieve it. Depending on the access structure, the number of multi-qubits may be larger than the size of $X$ and different members of the set $X$ may hold different numbers of qubits.  This enlarged number of qubits is denoted by $\overline{X}$. The following figure is used to denote such a sharing scheme, where, by $s_0$ we mean a quantum state (not a classical bit). \\
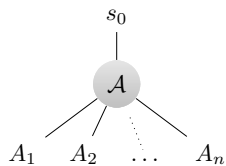
\begin{figure}[!htbp]
      \begin{center} 
  \scalebox{0.9}{
\begin{forest}
  styleB/.style={circle, top color=gray!20, bottom color=gray!50},
  [{$s_0$}
    [{$\cal{A}$}, styleB
      [{$A_1$}]
      [{$A_2$}]
      [{$\dots$}, edge={dotted, shorten <= 5pt}]
      [$A_n$]
    ]
  ]
\end{forest}}
\caption{This symbol means that the state $s_0$ can be recovered by the players $A_1$ to $A_n$ according to the access structure ${\cal A}$.}
\end{center}
\end{figure}
  \begin{remark}
    One can think of the bulb ${\cal A}$ both as the access structure and  as the encoding circuit which encodes the state $s_0$ to a multi-qubit state according to that structure. When such figures are concatenated, the corresponding quantum circuits are concatenated too. \\
  \end{remark}
  
Moreover, as we will show it is possible to do universal quantum computation on the same access structure. We will show that it is possible that a universal set of gates ${\cal P} = \{\cnot, H, S, T, X, Z \}$ be implemented on the shared state by local actions of each player on the qubit in his or her possession, eliminating  the overhead for encoding and decoding of the shared state and hence preventing any leakage of the input and output shared secret to the unauthorized parties. As we will see, all the above gates are implemented in a transversal way (share-wise gates) except the $T$ gate which requires communications between the players.

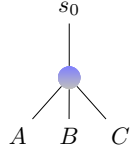
\begin{figure}[!htbp]
  \begin{center} 
    \scalebox{0.9}{
  \begin{forest}
    styleA/.style={circle, top color=blue!50, bottom color=gray!50},
    [{$s_0$}
      [ ,styleA
        [{$A$}]
        [{$B$}]
        [{$C$}]
      ]
    ]
  \end{forest}}
  \caption{The symbol with no sign on it always denote the (2,3) access structure. See the caption of figure (1).}
  \label{fig:simb23}
  \end{center}
\end{figure}

   
   We will frequently use the threshold access structure ${\cal A}:=(2,3)$ in which any two players can retrieve the state which has been shared between three players. This is denoted by Fig. \ref{fig:simb23}, the special figure where there is no symbol on the bulb:   \\

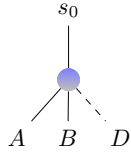
\begin{figure}[!htbp]
  \begin{center} 
    \scalebox{0.9}{
  \begin{forest}
    styleA/.style={circle, top color=blue!50, bottom color=gray!50},
    [{$s_0$}
      [ ,styleA
        [{$A$}]
        [{$B$}]
        [{$D$}, edge={dashed}]
      ]
    ]
  \end{forest}}
  \caption{ A dashed line means that the share of this player can be discarded. The other two players can still retrieve the state $s_0$ from the density matrix obtained by tracing over $D$.  }
  \label{fig:dashed23}
  \end{center}
\end{figure}

It may also happen that we have to discard some of the qubits in which case we denote the corresponding lines as dashed.   The necessity of discarding some of the qubits stems from the fact that  construction of non-maximal access structures can be achieved by discarding shares from secret sharing schemes with maximal access structure (Fig. \ref{fig:dashed23}).
 
This means that even if the pure state $|\overline{\psi}\ra$ is traced over $D$, the two parties $A$ and $B$ are still capable of retrieving the state $s_0$ from the remaining mixed state $\rho_{AB}$. 
More concretely if a state $|\psi\ra$ has been encoded to $|\overline{\psi}\ra_{ABD}$ such that any two players, say $A$ and $B$ can collaborate so that the initial state is recovered by one of them say A.  
This means that there is a recovery operation ${\cal R}_{_{AB}}$ such that
 \be {\cal R}_{_{AB}}(|\overline{\psi}\ra_{_{ABD}}\la \overline{\psi}|)=|\psi\ra_{_A}\la \psi|\otimes \chi_{_{BD}} \ee
Since $\Tr_{_D}$ commutes with ${\cal R}_{_{AB}}$, this means that the same kind of recovery operation by $A$ and $B$ will retrieve the state, that is:
\be{\cal R}_{_{AB}}\left(\Tr_{_{D}}(|\overline{\psi}\ra_{_{ABD}}\la \overline{\psi}|)\right)=|\psi\ra_{_A}\la \psi|\otimes \Tr_{_{D}}(\chi_{_{BD}})\ee

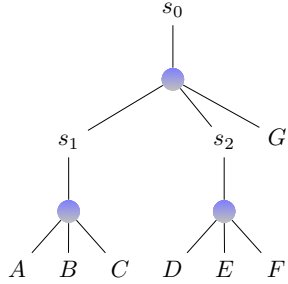
\begin{figure}[!htbp]
  \begin{center} 
    \scalebox{0.9}{
  \begin{forest}
    styleA/.style={circle, top color=blue!50, bottom color=gray!50},
    [{$s_0$}
      [ ,styleA
        [{$s_1$}
          [, styleA
            [$A$]
            [$B$]
            [$C$]
          ]
        ]
        [{$s_2$}
          [, styleA
            [$D$]
            [$E$]
            [$F$]
          ]
        ]
        [{$G$}]
      ]
    ]
  \end{forest}}
  \caption{Concatenated access structures: Depending on whether the extra share $G$ is given to $A$ or to $B$, different access structures are obtained.}
  \label{fig:concat}
  \end{center}
\end{figure}

 In such a case, the discarded share is represented by a dashed line. This action of discarding (tracing out) will play a major role in our construction of more complex concatenated schemes. 
\newpage
 Finally, we concatenate simple QSS schemes (expanding a share of access structure $S_1$ using access structure $S_2$) to implement more complex access structures \cite{QSS_GEN}. As an example consider Fig. \ref{fig:concat}. Depending on which of the participants in the list set $X=\{A, B, C, D, E, F\}$ will hold the share $G$, we can implement different access structures for the set  $X$, i.e. if $G=A$, then the access structure 
 contains the sets $AB$ and $AC$, while if $G=B$, it will contain the sets $AB$ and  $BC$. In both cases, the subsets of $\{D, E, F\}$ remain unauthorized. Note that $G$ being more than one qubit, can be shared between more than two members, i.e. it can be given to $A$ and $D$, in which case the access structure will be more complex. {Note that in concatenated schemes, no information can be gained from unauthorized sets of different instances of access structures \cite{QSS_GEN}. For example, in Fig. (\ref{fig:concat}), no information is leaked from $A$ and $D$ alone. Thus, while checking the authority of a given set, we have to be able to recover the secret recursively from the bottom if it is authorized, and it does not contain any information about the secret if it cannot recover the secret in this manner.}

\section{Quantum state sharing using 7-qubit code} \label{section3}
The original (2,3) scheme is based on using 3-level states as in (\ref{got}). We will construct all the QSS schemes from a basic threshold (2,3) scheme which is based on using the 7-qubit code. 
The 7-qubit code is a $[[7, 1, 3]]$ CSS code\cite{QEC_CSS}, which encodes one qubit into seven qubits in such a way that the distance between all the states involved is at least 3. The code is based on the classical $[7, 4, 3]$ hamming code and corrects one qubit error.  This code can be described by the following map:
\begin{align}
  \begin{split} \label{EQ:seven_qubit}
  \ket{0} \mapsto \ket{\overline{0}} = \frac{1}{\sqrt{8}}& \big(\ket{0000000} + \ket{1111000} + \ket{1100110} + \ket{1010101}\\
  +& \ket{0011110} + \ket{0101101} + \ket{0110011} + \ket{1001011}\big)\\
  \ket{1} \mapsto \ket{\overline{1}} = \frac{1}{\sqrt{8}}& \big(\ket{0000111} + \ket{1111111} + \ket{1100001} + \ket{1010010}\\
  +& \ket{0011001} + \ket{0101010} + \ket{0110100} + \ket{1001100}\big)\\
  \end{split}
\end{align}

From \cite{QSS_GEN}, we know that pure state erasure correcting codes are basically QSS schemes of maximal access structures. Now, let us distribute these 7 qubits among three different parties, ${A, B, C}$, and construct a $(2, 3)$ threshold scheme. Suppose that $A$ has qubits $\{1,2,3,4\}$, $B$ has $\{5\}$ and $C$ has $\{6, 7\}$. Since this is a pure QSS scheme, to prove its validity it suffices to prove that the density matrix of each unauthorized set is independent of the secret\cite{QSS_GEN, QEC_GEN}.
Assume that we are going to share $\ket{\psi_0} = \a \ket{0} + \b \ket{1}$. The shared secret can be described by state $\ket{\overline{\psi_0}}$, where $\ket{\overline{\psi_0}} = \a\ket{\overline{0}} + \b\ket{\overline{1}}$.\\
In this case, $\{A\}$, $\{B\}$ and $\{C\}$ are unauthorized. To compute the partial traces, it is useful to rewrite the logical qubits of Equation (\ref{EQ:seven_qubit}) as follows:

\begin{align}
  \begin{split}
  \ket{0} \mapsto \ket{\overline{0}} &= \frac{1}{2} \big(|G_{00}\ra |000\ra+ |G_{12}\ra |110\ra+ |G_{13}\ra |101\ra+ |G_{23}\ra |011\ra \big)\\
  \ket{1} \mapsto \ket{\overline{1}} &= \frac{1}{2} \big(|G_{00}\ra |111\ra+ |G_{12}\ra |001\ra+ |G_{13}\ra |010\ra+ |G_{23}\ra |100\ra \big)
  \end{split}
\end{align}
where $|G_{00}\ra$ is the four qubit GHZ state $\frac{1}{\sqrt{2}}(|0000\ra+|1111\ra)$
and $|G_{ij}\ra$ is obtained from $|G_{00}\ra$ by flipping the $i-$th and $j-$th qubit. \\

Thus, computing partial trace for every share produces the following density matrices which clearly are independent of the state $|\psi\ra$:
\begin{align}
  \begin{split}
  \rho_A = \Tr_{B, C} \big(\ketbra{\bar{\psi_0}} \big)  &= \frac{1}{4}\sum |G_{ij}\ra\la G_{ij}|,\quad \rho_B = \frac{1}{2} I_B,\quad \rho_C = \frac{1}{4} I_C\\
  \end{split}
\end{align}

Where $I_B$ is the identity matrix of size 2 (over $B$'s one qubit space) and $I_C$ is the identity matrix of size 4 (over $C$'s two qubit space). To obtain this results we used the condition that $|\a|^2+|\b|^2 = 1$. \\

Note that the authorized parties can recover the state, for  example $B$ and $C$ can recover the secret by computing the parity bit of their shares (applying two CNOTs from fifth and sixth qubit to the last qubit, and then applying two CNOTs from the last qubit to the fifth and sixth qubit). In fact with this sequence of actions, the last three qubits transform as
$
|i,j,k\ra\lo |j+k,i+k,i+j+k\ra
$
and hence
\begin{align}
  \begin{split}
  \ket{\overline{0}} &\mapsto |\xi\ra\otimes |0\ra\\
  \ket{\overline{1}} &\mapsto |\xi\ra\otimes |1\ra
  \end{split}
\end{align}
where
 \be
 |\xi\ra= \frac{1}{2} \big(|G_{00}\ra|00\ra+|G_{12}\ra|11\ra+|G_{13}\ra|01\ra+|G_{23}\ra|10\ra \big).
 \ee
  Hence the shared state $\a|\overline{0}\ra+\b|\overline{1}\ra$ transforms to $|\xi\ra\otimes (\a|0\ra+\b|1\ra)$ and is retrieved. We will now describe how concatenation of this scheme will lead to other more general schemes. 

\subsection{An $(n, n)$ QSS scheme} \label{subsection:newnn}

A number of $(n, n)$ QSS schemes have already been proposed in different contexts such as \cite{QSS_KN, HQSS_NN}. However most of these schemes use high-dimensional systems, which are not apt to the current schemes for actual quantum computation. Moreover, finding the right purification method for schemes that support universal quantum computation might not be straightforward. Thus, we are going to propose a new $(n, n)$ scheme that satisfies our requirements.\\ 

Consider the $(n,n)$ threshold scheme which obviously is not a maximal structure. Suppose the following hierarchy is applied to a secret state:

  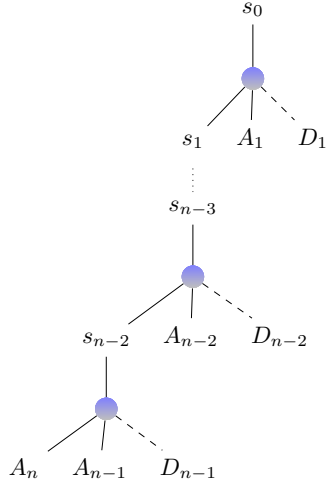
\begin{figure}[!htbp]
\begin{center} 
  \scalebox{0.9}{ 
\begin{forest}
  styleA/.style={circle, top color=blue!50, bottom color=gray!50},
  [{$s_0$}
    [, styleA
      [{$s_1$}
        [$s_{n-3}$, edge={dotted, shorten <= 5pt}
        [, styleA
          [$s_{n-2}$
            [, styleA
              [${A_n}$]
              [$A_{n-1}$]
              [$D_{n-1}$, edge={dashed}]
            ]
          ]
          [$A_{n-2}$]
          [$D_{n-2}$, edge={dashed}]
        ]
      ]
      ]
      [{$A_1$}]
      [{$D_1$}, edge={dashed}]
    ]
  ]
\end{forest}}
\caption{Hierarchical construction of the $(n,n)$ threshold scheme.}
\label{fig:n_n}
\end{center}
\end{figure}

Starting from the bottom of the Fig. \ref{fig:n_n}, we see that the players $A_{n-1}$ and $A_n$ can recover the state $s_{n-2}$ which with the information supplied by $A_{n-2}$ can lead to the recovery of the state in the upper level and so on until we reach the top of the figure where the collaboration of $A_1$ finally leads to the recovery of the encoded state. Note that in terms of quantum circuits, and in view of Equation (\ref{EQ:seven_qubit}) and the description following it on 7-qubit codes, a node like $s_1$ represents $4$ qubits and  the above figure implies that the 7-qubit code is applied to each one of the qubits in possession of $s_1$.  \\

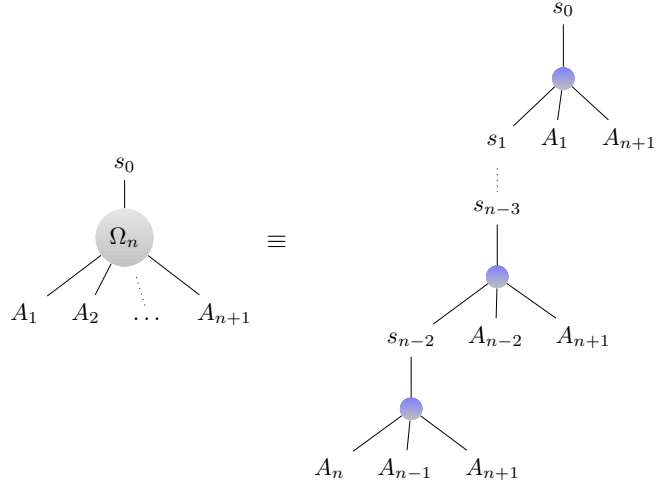
\begin{figure}[!htbp]
\begin{center}
\adjustbox{valign=c}{
  \scalebox{0.9}{ 
\begin{forest}
  styleB/.style={circle, top color=gray!20, bottom color=gray!50},
  [{$s_0$}
    [$\Omega_n$, styleB
      [$A_1$]
      [$A_2$]
      [$\dots$, edge={dotted, shorten <= 5pt}]
      [$A_{n+1}$]
    ]
  ]
\end{forest}}}
 $\equiv$
\adjustbox{valign=c}{
  \scalebox{0.9}{ 
\begin{forest}
  styleA/.style={circle, top color=blue!50, bottom color=gray!50},
  [{$s_0$}
    [, styleA
      [{$s_1$}
        [$s_{n-3}$, edge={dotted, shorten <= 5pt}
        [, styleA
          [$s_{n-2}$
            [, styleA
              [${A_n}$]
              [$A_{n-1}$]
              [$A_{n+1}$]
            ]
          ]
          [$A_{n-2}$]
          [$A_{n+1}$]
        ]
      ]
      ]
      [{$A_1$}]
      [{$A_{n+1}$}]
    ]
  ]
\end{forest}}} 
\caption{Hierarchical construction of the $\Omega_n$ access structure, eq. (\ref{Omegann}). }
\label{fig:A_nplus1}
\end{center}
\end{figure}

In this process, the shares $D_1$ to $D_{n-1}$ are discarded, that is they are traced over. This is a reflection of the non-maximality of the $(n,n)$ access structure and Corollary (2) of \cite{QSS_GEN}. Instead of discarding these shares, one can assemble them and give them to a new member $A_{n+1}$ according to the Fig. (\ref{fig:A_nplus1}). This will then correspond to a new access structure, denoted by $\Omega_n$ which will be explained in the next subsection. According to theorem (3) of \cite{QSS_GEN}, this scheme, being maximal, can be implemented by sharing a pure state.

\begin{remark}
One might ask why a scheme formed by concatenating simpler quantum secret sharing schemes is also a (secure) quantum secret sharing scheme and why unauthorized subsets in this new scheme contain no information about the secret. It is easy to prove in the case of maximal access structures since if $S\not\in\mathcal{A}(X)$ then we know $S^c\in\mathcal{A}(X)$ and $S$ contains no information about the secret. In addition, for non-maximal access structures, it is possible to apply the same argument on the purified version of that scheme. This is explicitly proved in the Gottesman's work On the Theory of Quantum Secret Sharing \cite{QSS_GEN}. 
\end{remark}

\subsection{A new access structure: The $\Omega_n$ scheme} \label{subsection:on}

The final building block that we need for constructing general access structures is a new and maximal access structure which we denote by $\Omega_n$ defined by its minimal authorized sets as 
 \be\label{Omegann}
  \Omega_n = \la A_1 A_2 \dots A_n, A_1 A_{n+1}, A_2 A_{n+1}, \dots ,A_n A_{n+1} \ra \ee
This means that any authorized set either contains $A_{n+1}$ or contains all other shares. We call $A_{n+1}$ the central share. Since this is a maximal access structure, from \cite{QSS_GEN}, a QSS scheme for it can be constructed by purifying an $(n, n)$ scheme. When we consider our previous construction of $(n, n)$ scheme, we achieved this construction by discarding $D_1, D_2, \dots, D_{n-1}$ from a pure state. Thus, if instead of discarding those shares, we produce a new share $A_{n+1}$, where $A_{n+1} = \{D_1, \dots, D_{n-1} \}$ we have effectively purified this scheme.\\

Same as before, $A_1 A_2 \dots A_n$ are able to recover the secret. In addition, $A_{n+1}$ can also recover the secret with the help of one of the other shares with the same method (recovering from $s_i$ to the top recursively).

\subsection {General schemes} \label{subsection:gen} 
Let us  start with a simple case. Assume that the set is $X=\{A, B, C\}$ and the access structure is ${\cal A}_0(X) = \la A B, A C \ra$. 
In the classical case, to share a secret string of bits $s$, one makes two copies of it and share it together with random strings $r_i$  according to the following scheme  \begin{align*}
  A_1 = s+r_1, \quad B_1 = r_2\\
  A_2 = s+r'_1, \quad C_1 = r'_2
\end{align*}
where $r_1+r_2=0$ and $r'_1+r'_2=0$.  
Here by $A_1$ and $A_2$ we mean the first and the second random string given to the player $A$. \\

\begin{figure}[!htbp]
\begin{center} 
  \scalebox{0.9}{
\begin{forest}
  styleB/.style={circle, top color=gray!20, bottom color=gray!50},
  [{$s_0$}
    [{$\Omega_2$}, styleB
      [{$s_1$}
        [{$\Omega_2$}, styleB
          [{$A_1$}]
          [{$B_1$}]
          [{$D_1$}, edge={dashed}]
        ]
      ]
      [{$s_2$}
        [{$\Omega_2$}, styleB
          [$A_2$]
          [$C_1$]
          [$D_2$, edge={dashed}]
        ]
      ]
      [{$s_3$}
        [{$A_3$}]
      ]
    ]
  ]
\end{forest}}
\caption{Hierarchical construction of the access structure ${\cal A}_0(X)=\la AB, AC\ra$. Note that $\Omega_2$ is the same as the (2,3) access structure shown in previous figures with a blue bulb. }
\label{fig:simple_ex}
\end{center}
\end{figure}
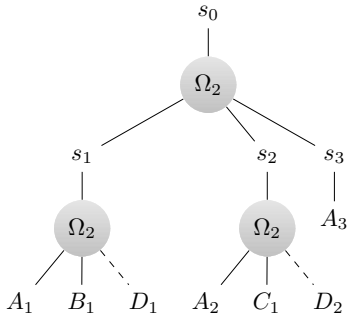

When we come to quantum state sharing, due to limitations from the no-cloning theorem \cite{NOCLON}, we have to  use a method similar to \cite{QSS_GEN} and  use the $\Omega_2$ scheme introduced in subsection \ref{subsection:on} as a substitute for copying. 
Fig. \ref{fig:simple_ex} is self-explanatory. Three shares are given to $A$, namely $A_1, A_2$ and $A_3$ and one share to  each of $B$ and $C$. The two shares $D_1$ and $D_2$ are redundant and are not used. 

\newpage
Starting from the bottom, $AB$ can retrieve $s_1$ and then with the share $A_3$ (in possession of $A$) retrieve $s_0$. A similar path exists also for $AC$, but none for $BC$. \\ 

To construct the scheme for any  access structure, we use induction. 
Assume that we already know how to construct all access structures with less than $n+1$ parties. Then we proceed with the following steps:\\

\begin{enumerate}[label=\textbf{Case \arabic*.}]
\item $\ {\cal A}_{n+1}$ is maximal:\\
In this case, we remove an arbitrary  player $x$ from ${\cal A}_{n+1}$ turning it into a non-maximal structure ${\cal A}_n$. Then by Theorem (3) of \cite{QSS_GEN}, the state which achieves the structure ${\cal A}_n$ between these players is necessarily mixed,  and any purification of this state has a unique ${\cal A}_{n+1}$ access structure. By purifying this mixed state and giving all the extra qubits which result from purification to the player $x$, we achieve a pure state sharing the state according to ${\cal A}_{n+1}$. \\
\end{enumerate}
\begin{example}
   Consider the set $X=\{A, B, C, E\}$ and the maximal structure $\overline{{\cal A}}=\la AE, BE, CE, ABC \ra$. To make a scheme for this, we remove $A$ and turning it to $\overline{{\cal A'}}=\la BE, CE\ra$. The scheme for this is already known and given by Fig. \ref{fig:simple_ex}, where $D_1$ and $D_2$ have been discarded. It is now enough to give these two shares to the removed player $A$. The final scheme is now given by Fig. \ref{fig:purification_ex}. 
   \begin{figure}[!htbp]
    \begin{center}
      \scalebox{0.9}{
      \begin{forest}
      styleB/.style={circle, top color=gray!20, bottom color=gray!50},
      [{$s_0$}
        [$\Omega_2$, styleB
          [$s_1$
            [$\Omega_2$, styleB
              [$E_1$]
              [$B_1$]
              [$A_1$, ]
            ]
          ]
          [$s_2$
            [$\Omega_2$, styleB
              [$E_2$]
              [$C_1$]
              [$A_2$]
            ]
          ]
          [$s_3$
            [$E_3$]
          ]
        ]
      ]
      \end{forest}}
      \caption{Constructing the access structure of example 2. }
      \label{fig:purification_ex}
    \end{center}
   \end{figure}
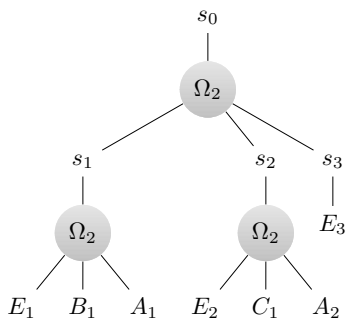

\end{example}

\begin{enumerate}[label=\textbf{Case \arabic*.}]
  \setcounter{enumi}{1}
\item $\ {\cal A}_{n+1}$ is not maximal:\\

In this case, it is obvious that the above method does not work, since if proceeded  as above, the final state would be pure which according to Corollary (2) of \cite{QSS_GEN} cannot correspond to ${\cal A}_{n+1}$ which is known to be a non-maximal access structure. Let this access structure be specified by its minimal authorized subsets $\la T_1, T_2, \cdots T_r \ra$ whose sizes are given by $|T_i|=k_i$. Obviously $|T_1\cup T_2\cup \cdots T_k|=n+1.$  Consider the Fig. \ref{fig:gen}.  Each of the states $s_i$ can be recovered by each group $T_i$. However, this by itself should not lead to the recovery of $s_0$ (otherwise no cloning will be violated). To remedy this, we expand
 $\la T_1, T_2, \cdots T_r \ra$ by adding subsets to it in order to make it maximal. Denote this expanded structure by $\overline{\tilde{{\cal A}}}_{n+1}$.  From case 1, we know how to share a secret $s_c$ to this structure. Now since every $T_i$ can recover its own secret $s_i$ and through membership in $\overline{\tilde{{\cal A}}}_{n+1}$
it can also recover the central share $s_c$, then by the property of $\Omega_r$, the secret $s_0$ can be recovered. \\

  \begin{figure}[!htbp]
\begin{center} 
  \scalebox{0.9}{
\begin{forest}
  styleB/.style={circle, top color=gray!20, bottom color=gray!50},
  [{$s_0$}
    [{$\Omega_r$}, styleB
      [{$s_1$}
        [{$\Omega_{k_1}$}, styleB
          [$\cross$ ]
          [{$\dots$}]
          [{$\cross$}]
          [$D_1$, edge={dashed}]
        ]
      ]
      [{$s_2$ } 
        [{$\Omega_{k_2}$}, styleB
          [{$\cross$}]
          [{$\dots$}]
          [{$\cross$}]
          [$D_2$, edge={dashed}]
        ]
      ]
      [$\dots$, edge={dotted, shorten <= 5pt}]
      [$\dots$, edge={dotted, shorten <= 5pt}]
      [$\dots$, edge={dotted, shorten <= 5pt}]
      [{$s_r$}
        [{$\Omega_{k_r}$}, styleB
          [{$\cross$}]
          [{$\dots$}]
          [{$\cross$}]
          [$D_r$, edge={dashed}]
        ]
      ]
      [{$s_c$}
        [$S'$, styleB
          [$A_1$]
          [$A_2$]
          [$\dots$]
          [$A_{n+1}$]
        ]
      ]
    ]
  ]
  \draw[thick, decorate, decoration={brace, amplitude=1.5em}]
  (-4.8, -4.5) --
  node[below=2em, font=\sffamily\bfseries]
    {$T_1$}
  (-7, -4.5)
  ;
  \draw[thick, decorate, decoration={brace, amplitude=1.5em}]
  (-1.3, -4.5) --
  node[below=2em, font=\sffamily\bfseries]
    {$T_2$}
  (-3.5, -4.5)
  ;
  \draw[thick, decorate, decoration={brace, amplitude=1.5em}]
  (2.4, -4.5) --
  node[below=2em, font=\sffamily\bfseries]
    {$T_r$}
  (0.2, -4.5)
  ;
\end{forest}}
\caption{The induction step for the case when ${\cal A}_{n+1}$ is not maximal. See the description in {\bf Case 2}}.
\label{fig:gen}
\end{center}
\end{figure}
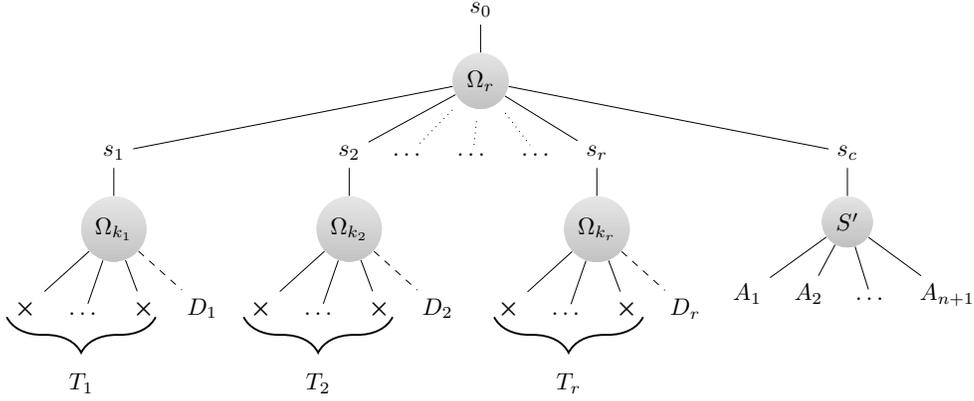
\end{enumerate}

\begin{remark} 
  The reader may ask why we have used the $\Omega_{k_i}$ schemes in Fig. \ref{fig:gen} to share $s_i$ to the set $T_i$,  while we could have also used any threshold scheme $(k_i\ ,\ k_i)$ for that purpose, for instance, the scheme that is proposed in \cite{HQSS_NN}, which also provides universal quantum computation. The reason is that the $\Omega_{k_i}$ schemes being maximal, lead to pure states and hence their concatenations will also be pure. Note that  all non-maximal access structures are produced by discarding some shares (i.e. $D_i$'s) from these pure concatenated schemes. Thus, in the purification step, one specific way of purification that also produces a concatenated 7-qubit code, is to  include the discarded $D_i$'s as a share of a new party member. This effectively purifies the non-maximal scheme.
\end{remark}

\begin{example} Consider again the set $X=\{A, B, C, E\}$ and the  non-maximal access structure ${\cal A}_4:={\cal A}(X) = \la ABC , BE, AE \ra$. This is not  maximal ( since neither $AB$ nor $CE$ are authorized). Therefore we follow the procedure of Fig. \ref{fig:gen_ex1}. The first step is to expand the secret using a $\Omega_3$ scheme, and distribute  the first three shares to the corresponding minimal authorized sets using $\Omega_n$ schemes:

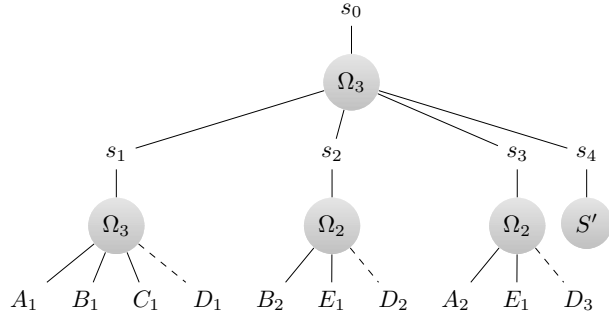
\begin{figure}[!htbp]
\begin{center} 
  \scalebox{0.9}{
\begin{forest}
  styleB/.style={circle, top color=gray!20, bottom color=gray!50},
  [{$s_0$}
    [{$\Omega_3$}, styleB
      [{$s_1$}
        [{$\Omega_3$}, styleB
          [{$A_1$}]
          [{$B_1$}]
          [{$C_1$}]
          [$D_1$, edge={dashed}]
        ]
      ]
      [{$s_2$}
        [{$\Omega_2$}, styleB
          [{$B_2$}]
          [{$E_1$}]
          [$D_2$, edge={dashed}]
        ]
      ]
      [{$s_3$}
        [{$\Omega_2$}, styleB
          [{$A_2$}]
          [{$E_1$}]
          [$D_3$, edge={dashed}]
        ]
      ]
      [{$s_4$}
        [$S'$, styleB]
      ]
    ]
  ]
\end{forest}}
\caption{Construction of the access structure in example 3. }
\label{fig:gen_ex1}
\end{center}
\end{figure}

We then amend ${\cal A}$ to a maximal structure $\overline{{\cal A}}=\la AE, BE, CE, ABC\ra$ for which we know how to implement a QSS from Fig. \ref{fig:purification_ex}. Putting these two figures together according to the general scheme Fig. \ref{fig:gen}, we obtain the scheme in Fig. \ref{fig:gen_ex2}.


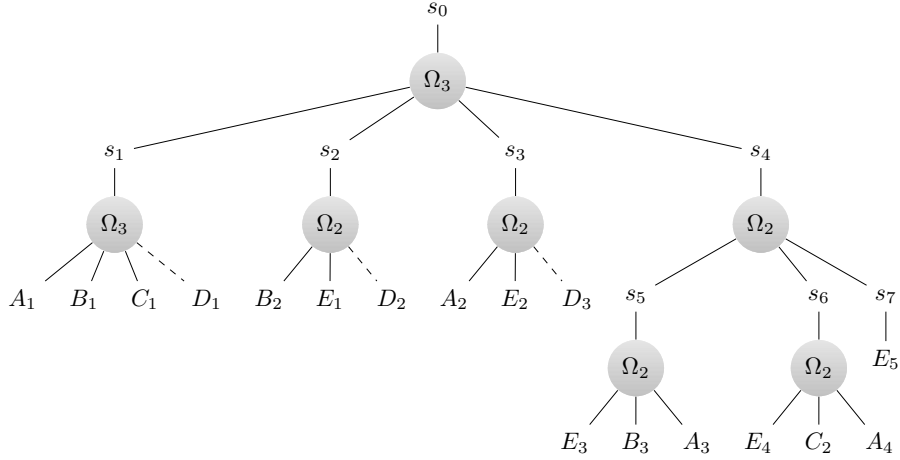
\begin{figure}[!htbp]
\begin{center} 
  \scalebox{0.9}{
\begin{forest}
  styleB/.style={circle, top color=gray!20, bottom color=gray!50},
  [{$s_0$}
    [{$\Omega_3$}, styleB
      [{$s_1$}
        [{$\Omega_3$}, styleB
          [{$A_1$}]
          [{$B_1$}]
          [{$C_1$}]
          [$D_1$, edge={dashed}]
        ]
      ]
      [{$s_2$}
        [{$\Omega_2$}, styleB
          [{$B_2$}]
          [{$E_1$}]
          [{$D_2$}, edge={dashed}]
        ]
      ]
      [{$s_3$}
        [{$\Omega_2$}, styleB
          [{$A_2$}]
          [{$E_2$}]
          [$D_3$, edge={dashed}]
        ]
      ]
      [{$s_4$}
        [$\Omega_2$, styleB
          [$s_5$
            [$\Omega_2$, styleB
              [$E_3$]
              [$B_3$]
              [$A_3$]
            ]
          ]
          [$s_6$
            [$\Omega_2$, styleB
              [$E_4$]
              [$C_2$]
              [$A_4$]
            ]
          ]
          [$s_7$
            [$E_5$]
          ]
        ]
      ]
    ]
  ]
\end{forest}}
\caption{Construction of the access structure in example 3.}
\label{fig:gen_ex2}
\end{center}
\end{figure}
The reader can now verify that every authorized set in $\la BE, AE, ABC\ra$ can retrieve the secret and none of the  unauthorized set can reach $s_0$. 
\end{example}

\section{Computing on shared secrets} \label{section4}

We have managed to share a qubit state according to any access structure among a set of players. The construction is all based on sharing a 7-qubit code in a concatenated scheme. To do universal quantum computation on these access structures, it is then enough to adopt the known techniques for doing quantum computation on 7-qubit codes. We begin with a description of universal gates on the 7-qubit code.

\subsection{Computing on the 7-qubit codes} \label{subsection:seven_qubit} 
It is well-known \cite{QEC_STAB} that the logical Pauli operators, the Hadamard and the CNOT gate can be implemented on the 7-qubit code, by their bit-wise transversal operation, that is: 
\be
\overline{X_a}={X_a}^{\otimes 7}, \h \overline{H}={H}^{\otimes 7}, \h \overline{\textrm{CNOT}}={\textrm{CNOT}}^{\otimes 7},
\ee
where $X_a$ is any Pauli operator and in the last relation, all the 7 bits of the first logical state are the control bit and all the 7 bits of the second logical state are the target bits. \\

Verification of relation for  the X and Z Pauli operators is simple and easily verified by looking at the structure of the logical qubit states $|\overline{0}\ra$ and $|\overline{1}\ra$ in Equation (\ref{EQ:seven_qubit}). The relation for Hadamard operator and CNOT is proved \cite{QEC_STAB} by noting that 
the stabilizers of the 7-qubit code and in fact any self-dual CSS code is either the product of $X$ operators or $Z$ operators in similar positions. In other words, the logical CNOT realized as $\overline{\textrm{CNOT}}=\textrm{CNOT}^{\otimes 7}$  has the same commutation relations with $\overline{X}$ and $\overline{Z}$ as the ordinary CNOT has with $X$ and $Z$.  \\ 

We also need to implement the  gate $S = \begin{pmatrix}1&0\\0&i \end{pmatrix}$ which transforms the eigenstates of the X operator to that of the Y operator. In view of the structure of the logical states $|\overline{1}\ra$ and  $|\overline{0}\ra$ in Equation (\ref{EQ:seven_qubit}) (i.e. the number of 1's in these states), it is obvious that we can implement the logical $\overline{S}$ gate as 

\be
\overline{S}={S^\dagger}^{\otimes 7}.
\ee

To make this set a universal set of gates, we  have to include the $\frac{\pi}{8}$ gate, $\overline{T}=\ketbra{\overline{0}} + e^{i\pi/4} \ketbra{\overline{1}}$. However, the problem is that this gate cannot be implemented directly and transversally as with the previous gates. To do this, we use gate teleportation \cite{GT_GOTT, GT_ZH} as shown in Fig. \ref{fig:gateteleport_oneQbit}, which explains the teleportation of the $T$ gate on one-qubit unencoded states.

\[
  \Qcircuit @C=.7em @R=.4em @! {
  \lstick{\ket{0}} & \gate{H} & \gate{T}  & \targ & \ctrl{1} &  \meter\\
  \lstick{\ket{\psi}} & \qw & \qw & \ctrl{-1} & \targ & \gate{SX} \cwx & \qw & \lstick{T\ket{\psi}}\\
  }
\]
\begingroup \vspace*{-\baselineskip}
\captionof{figure}{Gate teleportation of the $T$-gate on one-qubit.}
\label{fig:gateteleport_oneQbit}
\vspace*{\baselineskip}\endgroup

The state evolved through this circuit after the operation of the two CNOT gates is given by ( $\ket{\psi} = \a \ket{0} + \beta\ket{1}$):
\begin{align}
  \ket{0} \otimes \Big(\a \ket{0} + \b e^{i\pi/4}\ket{1}\Big) + \ket{1} \otimes \Big(\b \ket{0} + \a e^{i\pi/4}\ket{1}\Big).
\end{align}
Upon measuring the ancilla (first qubit) the second qubit projects either onto the state 
$T|\psi\ra$ or to a state which is corrected by the gate $SX$ to $T|\psi\ra$. In either case the gate $T$ is teleported by using the Hadamard gates, the CNOT and the $S$ and the $X$ gates.\\

We now upgrade  this circuit and adapt it to the present setting for implementation of the encoded $\overline{T}$ on logical states. (Fig. \ref{fig:gateteleport_NQbit})

\[
  \Qcircuit @C=.7em @R=1em @!{
  \lstick{\ket{0}} & \gate{H} & \gate{T} & \gate{{\cal E}} & \targ \qwmulti & \ctrl{1} \qwmulti & \meter \qwmulti\\
  \lstick{\ket{\overline{\psi}}} & \qwmulti & \qwmulti & \qwmulti & \ctrl{-1}\qwmulti & \targ \qwmulti  & \gate{\overline{S}\overline{X}} \qwmulti \cwx & \rstick{\overline{T}\ket{\overline{\psi}}} \qwmulti \\
  }
\]
\begingroup \vspace*{-\baselineskip}
\captionof{figure}{Gate teleportation of the encoded $\overline{T}$-gate on logical states}
\label{fig:gateteleport_NQbit}
\vspace*{\baselineskip}\endgroup

{We assume that a number of ancillary states are prepared in state $TH\ket{0}$, and are pre-shared among party members using the same QSS scheme}. The box ${\cal E}$ shows the encoding circuit which encodes qubit states $|0\ra$ and $|1\ra$ to logical states $|\overline{0}\ra$ and $|\overline{1}\ra$. The output of the circuit is now given by 

\begin{align}
  \ket{\overline{0}} \otimes \Big(\a \ket{\overline{0}} + \b e^{i\pi/4}\ket{\overline{1}}\Big) + \ket{\overline{1}} \otimes \Big(\b \ket{\overline{0}} + \a e^{i\pi/4}\ket{\overline{1}}\Big) \label{EQ:gen_t}
\end{align}

which shows that the output state is now given by $\overline{T}|\overline{\psi}\ra$ provided that we can do the correction $\overline{S}\overline{X}$ which we obviously can. The only problem is to see if by separable qubit-wise measurement we can determine the first logical qubit to be in $|\overline{0}\ra$ or $|\overline{1}\ra$. This is indeed possible by checking the parity of the 7-bits measured by the players as seen from Equation (\ref{EQ:seven_qubit}).  {Note that at some point in the hierarchy we might need to discard one of the shares of the $(2, 3)$ scheme constructed by the 7-qubit code. To make measurement possible based on the parity of bit-wise measurements, we need to discard the first four qubits while discarding one share. This also comes from Equation (\ref{EQ:seven_qubit}). Otherwise, the overall measurement result of bit-wise measurement is important for measuring the logical Z gate. For example, getting $00001$ for the first two shares means that we need to apply correction.} In this way, we have shown that by transversal bitwise gate operations and measurements, it is possible to do universal quantum computation on the 7-qubit code and hence do quantum computation on a $(2,3)$ scheme which is the basic building block of general access structures. Before proceeding to the computation, a note on security of the protocol, the difficulty of gaining information by unauthorized subsets,  is in order.  \\

\subsubsection{ A note on security } 
 First we should stress that  even in the field of classical cryptography, few protocols are proved to be information theoretically secure, rather most of them are proved to be practically secure in view of the large resources needed for their breaking.  In the present case, the assumption is that in the purified version of any scheme, which results in a maximal access structure, at least one authorized set is not dishonest. Furthermore, we assume that the shared state is completely secure after the sharing process, which means this proof will not work if the original state is tampered with, i.e. if is entangled with some qubits in possession of the dishonest party. Given this, let us consider a scenario in which a dishonest party tries to gain information about the secret by applying arbitrary operators and measurements during the computation. This would effectively disturb the information contained in its complement set \cite{QCRYPT}, leading to the revealing of the unauthorized leakage of information.  We prove this statement for maximal access structures, since any non-maximal access structure can be described by a maximal access structure with one share discarded \cite{QSS_GEN}. In maximal access structures, the complement of a dishonest party, which is not authorized, is an authorized set \cite{QSS_GEN}. In addition, from equations (\ref{EQ:seven_qubit} and \ref{EQ:gen_t}) we see that during the application of a $T$ gate, any information shared between these two group  is independent of the secret. The reason is that when written in the computational basis, the overall shared bits belong to the logical $0$ or logical $1$ sub-spaces each with probability $\frac{1}{2}$ and independent of their share. Therefore no additional information is leaked  to the dishonest party. At each step, the density matrix of the dishonest party remains independent of the secret since every operation is local and the authorized party should be able to recover the secret by following the protocol. This proves our claim.

\subsection{Computing on general access structures}
Generalizing  universal quantum computation on 7-qubit code to any access structure is now straightforward since the latter is made from the concatenation of the former. The problem in our context is  simpler than the one in \cite{QEC_STAB} which is devoted to fault-tolerant computation since in our case a basic step is to measure the logical Z operator as explained in subsection \ref{subsection:seven_qubit}, which need not be fault-tolerant. Assume that some n qubit secret $\ket{s}$ is shared among $n$ parties using a scheme that is implemented as explained in subsection \ref{subsection:gen}, with access structure S. Note that every qubit $i$ is shared independently. Furthermore, we use the same assumption as in \cite{HQSS_NN} for doing computation on QSS schemes. We assume that the desired circuits require less than $t$ number of $T$ gates in general. Thus, by pre-sharing $t$ number of $\tau=TH\ket{0}$ states, the encoding step is finished.\\

The next step is expanding our computation method while we expand a qubit in a hierarchy. We already know how to do transversal computation on 7-qubit code and a discarded 7-qubit code). As for any concatenated quantum code, each qubit has to apply a relevant gate according to its parent (the node above it in the diagram). Thus, while expanding a qubit using a 7-qubit code, it is obvious that $X, Z, \cnot, H$ can still be implemented transversally. However, $S$ gate and Z-measurement require more explanation. As for the $S$ gate, since $\overline{S} = {{S^\dagger}^{ \otimes 7}}$ and $\overline{S^\dagger} = S^{\otimes7}$ each new qubit can determine the appropriate gate based on its parent. Hence, qubits at an odd level have to apply the $S^\dagger$ gate, and qubits at an even level have to apply the $S$ gate. Furthermore, since measuring Z operator on an expanded ancillary qubit can be done by computing the parity bit of bit-wise measurements (if we use the last 3 qubits of 7-qubit code as a $(2,2)$ scheme) the overall parity bit can still determine whether there is a need for applying the correction $SX$ operator. Note that these measurements (even in the previous subsection) destroy any superposition of encoded ancillary state, which causes no problem in this context since these states have no use after the measurement.\\

Hence, the only modification needs to be done in the computation method for general access structures from the 7-qubit code is the application of the $S$ gate. Using this method, we are able to apply arbitrary quantum circuits with at most $t$ number of $T$ gates in their construction. However, as mentioned in \cite{HQSS_NN}, there is still the possibility that party members might be able to use a protocol to produce these shared ancillary $\tau$ states on demand.

\section{Conclusion and Outlook} \label{section5}
We proposed a method to construct QSS schemes for general access structures using only qubits, on which we are also able to do universal quantum computation with transversal opertions and pre-shared ancillary states. Our method only uses basic blocks of $(2,3)$-threshold scheme in contrast to more complicated $(k, n)$-threshold scheme mentioned in \cite{QSS_GEN}. The ease of computation is because our method is based on 7-qubit code, which also enables us to do experiments on more complex access structures with the current state of experimental quantum computation. These schemes might use exponential number of qubits depending on the access structure because of the purification steps. However, similar ideas might be used to construct usefull schemes such as threshold schemes more efficiently using concatenation of simple error correcting codes. It might also be possible to do a form of blind quantum computation using methods similar to other secure protocols \cite{HQSS_NN, FIT_DEL}
\vspace{-0.2cm}
\section{Acknowledegments} We thank the referees, specially one of them, whose very careful reading of the manuscript and very valuable comments led to a much better presentation of this article. This work was done with a support from the research council of the Sharif University of Technology, and with the financial support from Sharif University of Technology under grant no. G951418, and with partial support from Iran National Science Foundation under the grant INSF-96011347.
\vspace{-0.3cm}
\bibliographystyle{./unsrtabbrv}
\bibliography{./refs}

\begin{thebibliography}{10}

\bibitem{shamir1979share}
A.~Shamir.
\newblock How to share a secret.
\newblock {\em Communications of the ACM}, 22(11):612--613, 1979.

\bibitem{blakley1979safeguarding}
G.~R. Blakley et~al.
\newblock Safeguarding cryptographic keys.
\newblock In {\em Proceedings of the national computer conference}, volume~48,
  pages 313--317, 1979.

\bibitem{hillery1999quantum}
M.~Hillery, V.~Bu{\v{z}}ek, and A.~Berthiaume.
\newblock Quantum secret sharing.
\newblock {\em Physical Review A}, 59(3):1829, 1999.

\bibitem{karlsson1999quantum}
A.~Karlsson, M.~Koashi, and N.~Imoto.
\newblock Quantum entanglement for secret sharing and secret splitting.
\newblock {\em Physical Review A}, 59(1):162, 1999.

\bibitem{10}
A.~D. Smith.
\newblock Quantum secret sharing for general access structures.
\newblock {\em arXiv preprint quant-ph/0001087}, 2000.

\bibitem{QSS_KN}
R.~Cleve, D.~Gottesman, and H.-K. Lo.
\newblock How to share a quantum secret.
\newblock {\em Phys. Rev. Lett.}, 83:648--651, Jul 1999.

\bibitem{QSS_GEN}
D.~Gottesman.
\newblock Theory of quantum secret sharing.
\newblock {\em Phys. Rev. A}, 61:042311, Mar 2000.

\bibitem{imai2003quantum}
H.~Imai, J.~M{\"u}ller-Quade, A.~C. Nascimento, P.~Tuyls, and A.~Winter.
\newblock A quantum information theoretical model for quantum secret sharing
  schemes.
\newblock {\em arXiv preprint quant-ph/0311136}, 2003.

\bibitem{bai2016generalized}
C.-M. Bai, Z.-H. Li, T.-T. Xu, and Y.-M. Li.
\newblock A generalized information theoretical model for quantum secret
  sharing.
\newblock {\em International Journal of Theoretical Physics},
  55(11):4972--4986, 2016.

\bibitem{1}
C.~H. Bennett, F.~Bessette, G.~Brassard, L.~Salvail, and J.~Smolin.
\newblock Experimental quantum cryptography.
\newblock {\em Journal of Cryptology}, 5(1):3--28, Jan 1992.

\bibitem{2}
S.~Gröblacher, T.~Jennewein, A.~Vaziri, G.~Weihs, and A.~Zeilinger.
\newblock Experimental quantum cryptography with qutrits.
\newblock {\em New Journal of Physics}, 8(5):75, 2006.

\bibitem{3}
W.-Y. Liang, M.~Li, Z.-Q. Yin, W.~Chen, S.~Wang, X.-B. An, G.-C. Guo, and Z.-F.
  Han.
\newblock Simple implementation of quantum key distribution based on
  single-photon bell-state measurement.
\newblock {\em Phys. Rev. A}, 92:012319, Jul 2015.

\bibitem{4}
A.~Broadbent, J.~Fitzsimons, and E.~Kashefi.
\newblock Proceedings of the 50th annual ieee symposium on foundations of
  computer science.
\newblock 2009.

\bibitem{5}
S.~Barz, E.~Kashefi, A.~Broadbent, J.~F. Fitzsimons, A.~Zeilinger, and
  P.~Walther.
\newblock Demonstration of blind quantum computing.
\newblock {\em Science}, 335(6066):303--308, 2012.

\bibitem{6}
Y.~Ouyang, S.-H. Tan, and J.~F. Fitzsimons.
\newblock Quantum homomorphic encryption from quantum codes.
\newblock {\em Phys. Rev. A}, 98:042334, Oct 2018.

\bibitem{7}
V.~Karimipour and M.~Asoudeh.
\newblock Quantum secret sharing and random hopping: Using single states
  instead of entanglement.
\newblock {\em Phys. Rev. A}, 92:030301, Sep 2015.

\bibitem{8}
S.~Bagherinezhad and V.~Karimipour.
\newblock Quantum secret sharing based on reusable greenberger-horne-zeilinger
  states as secure carriers.
\newblock {\em Phys. Rev. A}, 67:044302, Apr 2003.

\bibitem{9}
X.-L. Song, Y.-B. Liu, H.-Y. Deng, and Y.-G. Xiao.
\newblock (t, n) threshold d-level quantum secret sharing.
\newblock {\em Scientific Reports}, 7(1):6366, 2017.

\bibitem{gordon2006generalized}
G.~Gordon and G.~Rigolin.
\newblock Generalized quantum-state sharing.
\newblock {\em Physical Review A}, 73(6):062316, 2006.

\bibitem{qin2015proactive}
H.~Qin and Y.~Dai.
\newblock Proactive quantum secret sharing.
\newblock {\em Quantum Information Processing}, 14(11):4237--4244, 2015.

\bibitem{HQSS_NN}
Y.~Ouyang, S.-H. Tan, L.~Zhao, and J.~F. Fitzsimons.
\newblock Computing on quantum shared secrets.
\newblock {\em Physical Review A}, 96(5):052333, 2017.

\bibitem{fortescue2012reducing}
B.~Fortescue and G.~Gour.
\newblock Reducing the quantum communication cost of quantum secret sharing.
\newblock {\em IEEE Transactions on Information Theory}, 58(10):6659--6666,
  2012.

\bibitem{bai2017quantum}
C.-M. Bai, Z.-H. Li, M.-M. Si, and Y.-M. Li.
\newblock Quantum secret sharing for a general quantum access structure.
\newblock {\em The European Physical Journal D}, 71(10):255, 2017.

\bibitem{QEC_GEN}
E.~Knill, R.~Laflamme, and L.~Viola.
\newblock Theory of quantum error correction for general noise.
\newblock {\em Phys. Rev. Lett.}, 84:2525--2528, Mar 2000.

\bibitem{QEC_STAB}
D.~Gottesman.
\newblock Stabilizer codes and quantum error correction.
\newblock {\em arXiv preprint quant-ph/9705052}, 1997.

\bibitem{NOCLON}
W.~K. Wootters and W.~H. Zurek.
\newblock A single quantum cannot be cloned.
\newblock {\em Nature}, 299(5886):802--803, 1982.

\bibitem{QEC_CSS}
A.~R. Calderbank and P.~W. Shor.
\newblock Good quantum error-correcting codes exist.
\newblock {\em Phys. Rev. A}, 54:1098--1105, Aug 1996.

\bibitem{GT_GOTT}
D.~Gottesman and I.~L. Chuang.
\newblock Demonstrating the viability of universal quantum computation using
  teleportation and single-qubit operations.
\newblock {\em Nature}, 402(6760):390, 1999.

\bibitem{GT_ZH}
X.~Zhou, D.~W. Leung, and I.~L. Chuang.
\newblock Methodology for quantum logic gate construction.
\newblock {\em Phys. Rev. A}, 62:052316, Oct 2000.

\bibitem{QCRYPT}
C.~H. Bennett, G.~Brassard, and N.~D. Mermin.
\newblock Quantum cryptography without bell's theorem.
\newblock {\em Phys. Rev. Lett.}, 68:557--559, Feb 1992.

\bibitem{FIT_DEL}
V.~Dunjko, J.~F. Fitzsimons, C.~Portmann, and R.~Renner.
\newblock Composable security of delegated quantum computation.
\newblock In {\em International Conference on the Theory and Application of
  Cryptology and Information Security}, pages 406--425. Springer, 2014.

\end{thebibliography}


\begin{thebibliography}{}
\bibitem{1} Bennett, C.H., Bessette, F., Brassard, G. et al. J. Cryptology (1992) 5: 3. https://doi.org/10.1007/BF00191318
\bibitem{2} Simon Groeblacher, Thomas Jennewein, Alipasha Vaziri, Gregor Weihs, Anton Zeilinger,	New J. Phys. 8 (2006)
\bibitem{3} Wen-Ye Liang, Mo Li, Zhen-Qiang Yin, Wei Chen, Shuang Wang, Xue-Bi An, Guang-Can Guo, and Zheng-Fu Han
Phys. Rev. A 92, 012319 (2015)
\bibitem{4} Anne Broadbent, Joseph Fitzsimons, Elham Kashefi, Proceedings of the 50th Annual IEEE Symposium on Foundations of Computer Science (FOCS 2009), pp. 517-526
\bibitem{5} Stefanie Barz, Elham Kashefi, Anne Broadbent, Joseph F. Fitzsimons, Anton Zeilinger, Philip Walther, Science  20 Jan 2012: Vol. 335, Issue 6066, pp. 303-308 DOI: 10.1126/science.1214707
\bibitem{6} Y. Ouyang, S.-H. Tan, and J. Fitzsimons. Quantum homomorphic encryption from quantum codes, August 2015. arXiv:1508.00938.
\bibitem{7} V. Karimipour and M. Asoudeh, Phys. Rev. A 92, 030301(R) (2015)
\bibitem{8} Saber Bagherinezhad and Vahid Karimipour
Phys. Rev. A 67, 044302 (2003)
\bibitem{9} Xiu-Li Song, Yan-Bing Liu, Hong-Yao Deng, Yong-Gang Xiao, Scientific Reports 7, Article number: 6366 (2017) doi:10.1038/s41598-017-06486-4

\bibitem{10} Adam D. Smith, arXiv preprint quant-ph/0001087, (2000)

\bibitem{QSS_KN} Richard Cleve, Daniel Gottesman, and Hoi-Kwong Lo Phys. Rev. Lett. 83, 648 (1999)
\bibitem{QSS_GEN} Daniel Gottesman Phys. Rev. A 61, 042311 (2000)
\bibitem{FQC_POLY} 	D. Aharonov, M. Ben-Or, Proceeding STOC '97 Proceedings of the twenty-ninth annual ACM symposium on Theory of computing (1997)
\bibitem{QEC_CSS} A. R. Calderbank, Peter W. Shor, Phys. Rev. A 54, 1098 (1996)
\bibitem{QEC_STAB} Daniel Gottesman, 	preprint arXiv:quant-ph/9705052 (1997)
\bibitem{NOCLON} W. K. Wootters, W. H. Zurek, Nature 299, 802â803 (1982)
\bibitem{HQSS_NN} Yingkai Ouyang, Si-Hui Tan, Liming Zhao, and Joseph F. Fitzsimons
Phys. Rev. A 96, 052333 (2017)
\bibitem{FQC_DDIM} Gottesman D. (1999) Fault-Tolerant Quantum Computation with Higher-Dimensional Systems. In: Williams C.P. (eds) Quantum Computing and Quantum Communications. Lecture Notes in Computer Science, vol 1509. Springer, Berlin, Heidelberg
\bibitem{QEC_GEN} Emanuel Knill, Raymond Laflamme, and Lorenza Viola
Phys. Rev. Lett. 84, 2525 (2000)

\bibitem{QCRYPT} Charles H. Bennett, Gilles Brassard, and N. David Mermin, Phys. Rev. Lett. 68, 557(1992)
\bibitem{GT_GOTT} Daniel Gottesman, Isaac L. Chuang, Nature 402, 390â393 (1999)
\bibitem{GT_ZH} Xinlan Zhou, Debbie W. Leung, Isaac L. Chuang, Phys. Rev. A 62, 052316 (2000)

\end{thebibliography}

\end{document}